\documentclass[aip,pop,preprint,showpacs]{revtex4-1}
\usepackage{amsmath}
\usepackage{amssymb}
\usepackage{color}
\usepackage{epsfig}
\usepackage{graphicx}
\usepackage{bm}
\usepackage{url}

\begin{document}
\title{Three-dimensional simulations of the orientation and structure
of reconnection X-lines}

\author{R. Schreier}
\author{M. Swisdak} 
\email{swisdak@umd.edu}
\author{J. F. Drake}
\affiliation{IREAP, University of Maryland, College Park, MD
20742-3511, USA}
\author{P. A. Cassak}
\affiliation{Department of Physics, West Virginia University,
Morgantown, West Virginia 26506, USA}

\date{\today}

\begin{abstract}

This work employs Hall magnetohydrodynamic (MHD) simulations to study
the X-lines formed during the reconnection of magnetic fields with
differing strengths and orientations embedded in plasmas of differing
densities.  Although random initial perturbations trigger the growth
of X-lines with many orientations, at late time a few robust X-lines
sharing an orientation reasonably consistent with the direction that
maximizes the outflow speed, as predicted by Swisdak and Drake
[Geophys. Res. Lett., {\bf 34}, L11106, (2007)], dominate the system.
The existence of reconnection in the geometry examined here
contradicts the suggestion of Sonnerup [J. Geophys. Res. {\bf 79},
1546 (1974)] that reconnection occurs in a plane normal to the
equilibrium current.  At late time the growth of the X-lines
stagnates, leaving them shorter than the simulation domain.
\end{abstract}

\pacs{52.35.Vd, 94.30.cp, 96.60.lv, 52.65.Kj}

\maketitle


According to the frozen-in theorem of MHD, two adjoining collisionless
plasmas with different densities, temperatures, and magnetic fields
cannot alter their magnetic topology, and hence transport across their
common boundary is prohibited. Magnetic reconnection violates this
constraint as, for example, when solar wind plasma penetrates the
Earth's magnetosphere or fusion plasma escapes from a tokamak core
during a disruption.  The questions of whether and how reconnection
takes place for arbitrary plasma conditions are important for these
and other systems.

\begin{figure}
\begin{center}
\noindent\includegraphics[width=\textwidth]{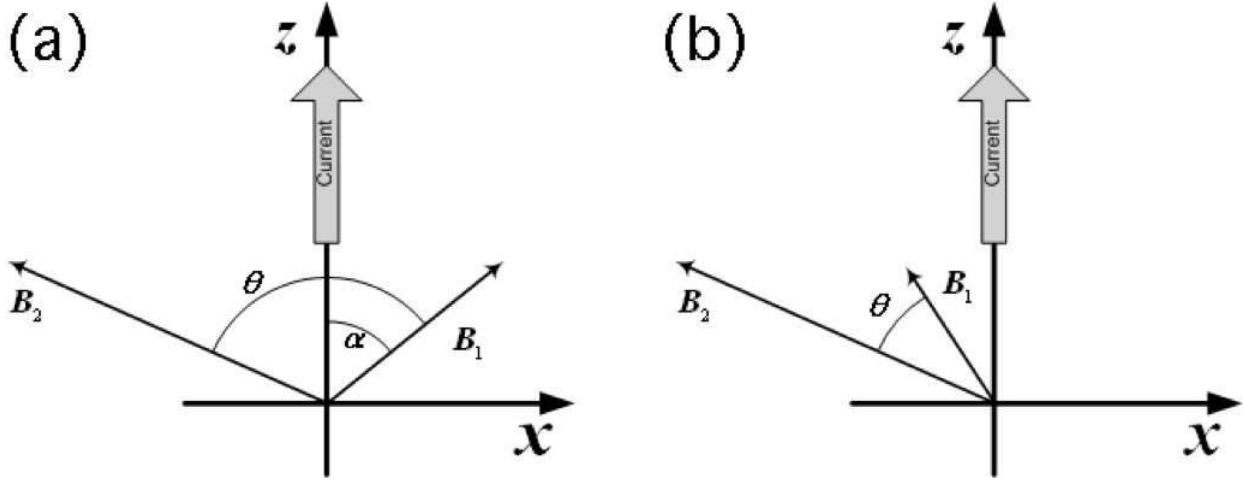}
\caption{\label{Illustration} Field line geometries related to the
\citet{sonnerup74a} hypothesis.  The reconnecting plasmas with fields
$B_1$ and $B_2$ occupy the spaces $y>0$ and $y<0$.  In both panels the
X-line parallels $\hat{z}$, reconnection occurs in the $x-y$ plane,
and the fields are oriented such that the components parallel to the
X-line are equal.  \citet{sonnerup74a} proposed that reconnection
occurs when the $x$-components of $B_1$ and $B_2$ are anti-parallel
(a), and that it otherwise does not (b).}
\end{center}
\end{figure}

Consider two plasmas threaded by magnetic fields ${\bf B}_{1}$ and
${\bf B}_{2}$ of arbitrary relative orientation and separated by a
planar surface, the $x-z$ plane in Fig.~\ref{Illustration}.  We define
the $y$ axis to be perpendicular to the discontinuity plane, the $z$
axis to parallel the X-line, and the $x$ axis to complete the
right-handed triplet.  Let $\theta$ be the shear angle between the two
fields and $\alpha$ be the unknown angle between ${\bf B}_{1}$ and the
X-line.  In the highly symmetric cases often considered in theory and
simulations $\alpha$ can frequently be easily deduced (e.g., $\alpha =
90^{\circ}$ for $\mathbf{B_1}=-\mathbf{B_2}$,).  For more general
configurations, however, no obvious choice exists, nor is it even
clear that a single X-line orientation will dominate the
system. \citet{sonnerup74a} argued that $\alpha$ is determined by
requiring that the currents in the reconnection plane vanish or,
equivalently, that the components of the fields parallel to the X-line
(the guide fields) be equal.  As a consequence, no reconnection occurs
in this scenario when $\cos \theta \geq B_1/B_2$ (assuming, as in
Fig.~\ref{Illustration}, $B_1\leq B_2$), since no component of the
field changes sign across the discontinuity.

Others \cite{cowley76a,teh08a} have questioned this choice on both
theoretical and observational grounds.  For example, although the
Sonnerup criterion implies that reconnection between plasmas with
small shear angles occurs infrequently, in situ observations in the
solar wind suggest the contrary to be true
\citep{gosling07a,gosling07b,phan09a,phan10a}.  In fact, most
reconnection events in the solar wind occur at shear angles
$<90^{\circ}$ \citep{gosling07b,phan10a}.

As an alternative, \citet{swisdak07a} proposed that the X-line orients
itself so as to maximize the speed of the Alfv\'{e}nic outflow.  The
outflow speed for plasmas with reconnecting components $B_{1x}$ and
$B_{2x}$ and mass densities $\rho_{1}$ and $\rho_{2}$ is
\citep{swisdak07a,cassak07a}
\begin{equation}
\label{vout}
v_{\text{out}}^{2} = \frac{B_{1x} + B_{2x}}{4 \pi} \left(
\frac{\rho_{1}}{B_{1x}} + \frac{\rho_{2}}{B_{2x}} \right)^{-1}.
\end{equation}
Writing this expression in terms of $\alpha$ and maximizing with
respect to $\alpha$ for a fixed $\theta$ determines the X-line
orientation.  Since $v_{\text{out}}$ always has a local maximum
between $\alpha=0$ and $\alpha=\theta$ reconnection occurs for any
$\theta \neq 0$.  An alternative suggestion holds that maximizing a
related quantity, the normalized reconnection rate, determines the
X-line orientation (M.~A.~Shay, private communication).

In this work we perform 2-D and 3-D two-fluid simulations of
reconnection between asymmetric plasmas in order to explore the
generic development of X-lines.  We first use a 2-D simulation to
demonstrate magnetic reconnection in a system with small shear angle
with the caveat that, since 2-D simulations artificially impose the
orientation of the X-line, studying the full development of the system
necessitates a 3-D domain.  Hence, we also consider a 3-D simulation
of the same system.  In previous investigations of 3D Hall
reconnection \citep{huba02b,shay03a,lapenta06a}, the initial
configuration of anti-parallel fields confined nascent X-lines to one
plane between the two plasmas.  Initially localized X-lines grew in
the direction of the electron current and, in some cases, extended
over almost the entire computational domain.  For the more general
situation considered here, X-lines in the linear stage of development
grow on different planes, known as rational surfaces, and undergo more
complex interactions.  We find that X-lines of several different
orientations are excited at early times, but eventually only a few
modes dominate.  Interestingly, and in contrast to previous
investigations \citep{huba02b,shay03a}, the X-lines' length stagnates
at a finite value that is shorter than the simulation domain.


For our initial equilibrium we employ a double tearing mode
configuration with magnetic field components
\begin{equation}
B_x(y) = \tanh\left(\frac{y + L_y/4}{w_0}\right) - \tanh\left(\frac{y
- L_y/4}{w_0}\right) - 1 \label{bxeq}
\end{equation}
and
\begin{equation}\label{bzeq}
B_z(y) =
- \frac{1}{\sqrt{2}}\tanh\left(\frac{y + L_y/4}{w_0}\right)
+ \frac{1}{\sqrt{2}}\tanh\left(\frac{y - L_y/4}{w_0}\right) 
  + 2\sqrt{2},
\end{equation}
where $w_0=0.5$ is the initial width of the current sheet (the
normalization is described later).  The asymptotic fields have
components $(B_x,B_y,B_z)=(1,0, \sqrt{2})$ and $(-1,0,2\sqrt{2})$.
Total pressure is balanced using a non-uniform number density $n$
given by
\begin{equation}\label{ne}
n = \frac{1}{T}\left(P_a - \frac{B^2}{2}\right),
\end{equation}
where the temperature $T = 1.0$ is uniform, $B$ is the magnitude of
the magnetic field, $P_a = 5.5$ is a constant, and the factor of 2 in
the denominator arises from our code's normalization.  For this
configuration $\rho_1 = 4$, $\rho_2=1$, $B_{1} = \sqrt{3}$, $B_{2} =
3$, and the shear angle $\theta = 54.7^{\circ}$.  This system
represents the limiting case described by \citet{sonnerup74a} since
$\cos \theta = B_{1} / B_{2}$.

The numerical simulations use the Hall-MHD code F3D
\citep{shay04a}.  It explicitly advances the dynamical variables
(magnetic field, mass density, and ion velocity) with the second-order
trapezoidal leapfrog method \citep{guzdar93a} in time and fourth-order
finite differencing in space.  Periodic boundary conditions are
applied in all directions.  Variables are measured in normalized
units: lengths to the ion inertial length $d_i=(m_ic^2/4\pi
n_0e^2)^{1/2}$, velocities to the Alfv\'{e}n speed $c_A=B_0/(4\pi m_i
n_0)^{1/2}$, densities to an arbitrary value $n_0$, pressures to
$P_0=m_i n_0c_A^2$, magnetic fields to an arbitrary field strength
$B_0$, temperatures to $m_i c_A^2$ and electric fields to
$E_0=c_AB_0/c$. Here $c$ is the speed of light, and $m_i$ and $e$ are
the mass and charge of the ions, respectively.

The grid cells have a length of $0.2$ on each side.  No explicit
viscosity or resistivity are applied, but a fourth-order diffusion
coefficient of $10^{-3}$ damps noise at the grid scale.  The
electron-to-ion mass ratio is $m_e/m_i=1/25$.  Since the electron
inertial length $d_e = d_i\sqrt{m_e/m_i}$ equals the cell size, the
simulations do not describe the details of the electron dynamics.


To determine whether reconnection can occur between the fields of
Eqs.~(\ref{bxeq}) and (\ref{bzeq}), we first perform a 2-D simulation.
The computational domain has size $L_x\times L_y=51.2\times 25.6$ and
no variations are allowed in the $z$ direction (i.e.,
$\partial/\partial z=0$).  We initiate reconnection with random
magnetic perturbations of amplitude $10^{-3} B_0$.  The perturbations
are generated in $k$-space with maximum wavenumbers of $k_x=k_y=15$.
Setting the initial perturbation amplitude 100 times smaller produces
similar final results.

\begin{figure}
\begin{center}
\noindent\includegraphics[width=\textwidth]{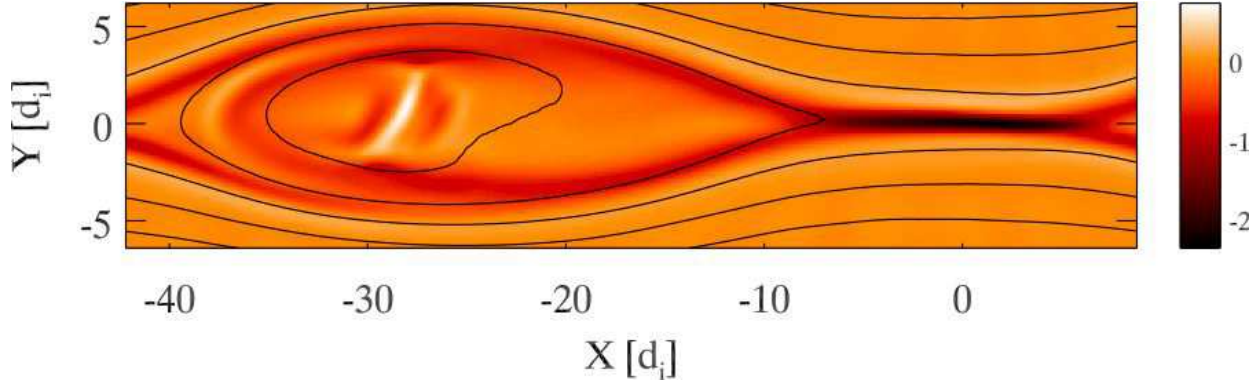} 
\caption{\label{2D_image} (Color online) Out-of-plane current density
$J_z$ for the lower sheet in the 2-D simulation at $t=156$ with
over-plotted magnetic field lines.  The other sheet exhibits similar
behavior.}
\end{center}
\end{figure}

During the early stages of the simulation, strong out-of-plane
currents develop, indicating the existence of magnetic reconnection.
As the system evolves, multiple X-lines form, move along the $x$-axis,
and merge, eventually leaving just one reconnection site.  The
out-of-plane current density of the lower current sheet after this
merging is shown in Fig.~\ref{2D_image}.  Following some initial
fluctuations, caused by the interactions of multiple X-lines, the
normalized reconnection rate stabilizes at a relatively steady value
of $\sim 0.02$.

This simulation demonstrates that reconnection can occur in a plane
that includes equilibrium currents in a geometry where the model of
\citet{sonnerup74a} suggests it should not.  However, since the
geometry of the computational domain determines the X-line orientation
it did not establish whether an optimal orientation exists.  Doing
that requires a full 3D system, in which X-lines are free to develop
in any direction.  Our 3D simulation uses the same initial equilibrium
as the 2D run, but applied on a computational domain of size
$L_x\times L_y\times L_z=51.2\times 51.2\times 409.6$, and a grid
scale in the out-of-plane direction of 0.2.  Initial perturbations on
the magnetic field in the $z$ direction have a maximum wavenumber of
$k_z=5$.

In the linear theory of the tearing mode in periodic systems,
reconnection can only occur at discrete locations, called rational
surfaces, where ${\mathbf k} \boldsymbol{\cdot} {\mathbf B}_{0} = 0$
(${\mathbf k}$ is the wavenumber of the linear mode and ${\mathbf
B}_{0}$ is the equilibrium field).  Due to the periodicity of the
domain, the wavenumbers of the linear instability must take the form
$k_{x} = 2 \pi m / L_{x}$ and $k_{z} = 2 \pi n / L_{z}$, where $m$ and
$n$ are integers.  This establishes rational surfaces at the locations
satisfying
\begin{equation}\label{rational_surfaces}
\frac{B_x(y)}{B_z(y)} = -\frac{nL_x}{mL_z}.
\end{equation}
Since $L_z/L_x=8$ several linear modes with $n\ne0$ can grow in the
current layer.  Once they exit the linear regime interactions between
different modes allow them to no longer respect the rational surfaces

To reiterate, the model of \citet{sonnerup74a} predicts that
reconnection will not occur in this system.  In contrast,
\citet{swisdak07a} predict that reconnection will occur with the
X-line at an angle $\alpha$ given by the root 
of the equation
\begin{eqnarray}\label{alpha}
0 = &&\frac{\rho_2}{\rho_1} \sin^2\alpha \left[\sin(\theta - 2\alpha)
- \frac{B_2}{B_1}\sin(2\theta - 2\alpha )\right] \nonumber \\ &+&
\frac{B_2}{B_1} \sin^2(\theta - \alpha) \left[\sin 2\alpha +
\frac{B_2}{B_1}\sin(\theta - 2\alpha )\right]
\end{eqnarray}
lying between $0$ and $\theta$.  Numerically solving for the
parameters of the present system --- $B_{2} / B_{1} =
\sqrt{3}, \rho_{2} / \rho_{1} = 1 / 4$, and $\theta = 54.7^{\circ}$ ---
yields $\alpha = 34.3^{\circ}$.

\begin{figure}
\begin{center}
\noindent\includegraphics[width=\textwidth]{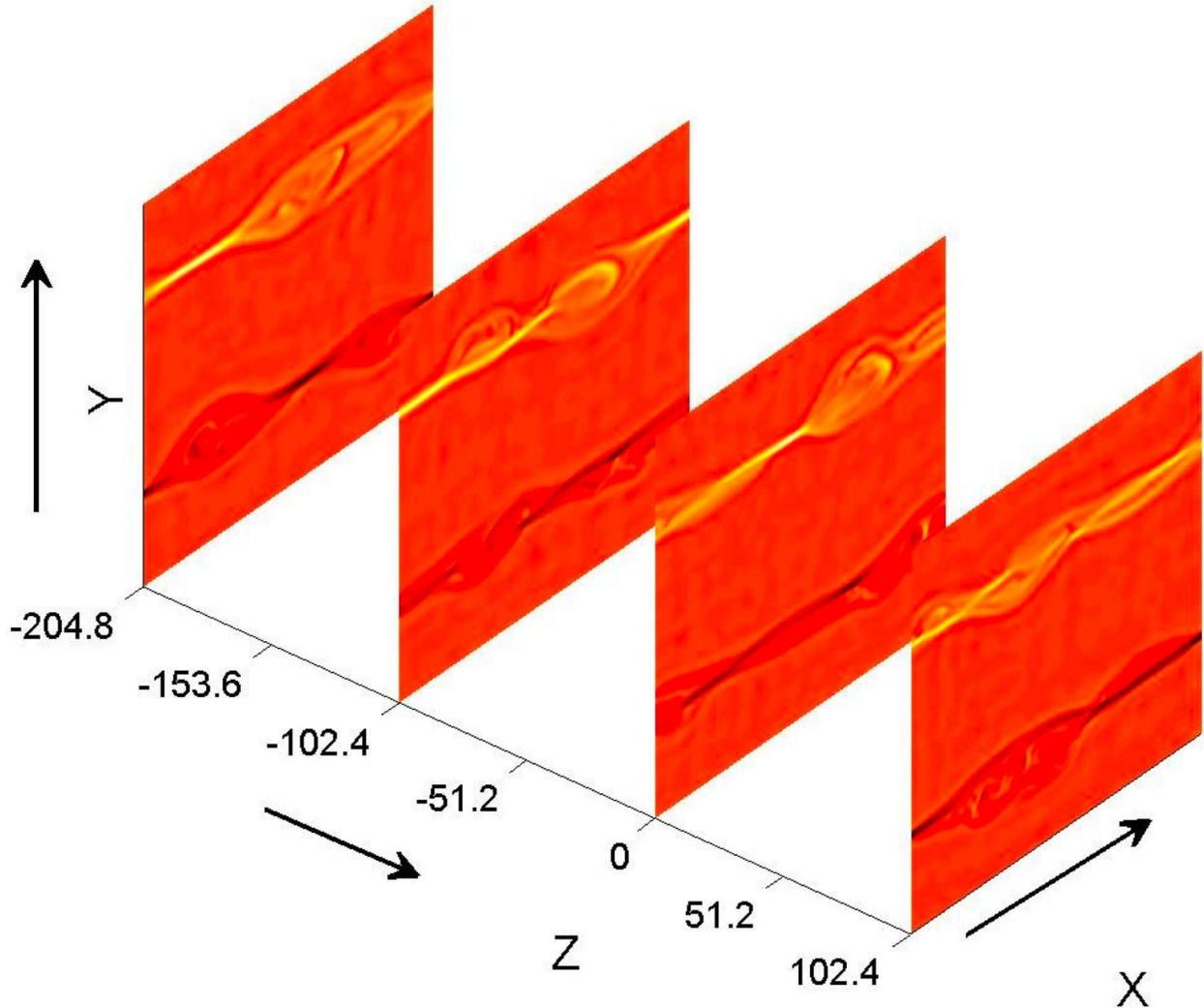}
\caption{\label{slices} (Color online) Slices of $J_{z}$ in the $x-y$
plane of the 3-D domain at $t=84$.  The upper (lower) current sheet is
described by bright (dark) colors, representing positive (negative)
currents.}
\end{center}
\end{figure}

The early evolution of the 3-D simulation mirrors that of the 2-D run,
in that both current sheets develop multiple X-lines separated by
bulges in which reconnected flux accumulates.  In Fig.~\ref{slices} we
present cuts at $t=84$ (relatively early in the simulation) of $J_z$
in the $x$-$y$ plane for four values of $z$ separated by $L_z/4$.  For
the 2-D simulation of Fig.~\ref{2D_image} the X-lines necessarily
parallel the $z$-axis, but that is not the case in Fig.~\ref{slices}.

\begin{figure}
\begin{center}
\noindent\includegraphics[width=\textwidth]{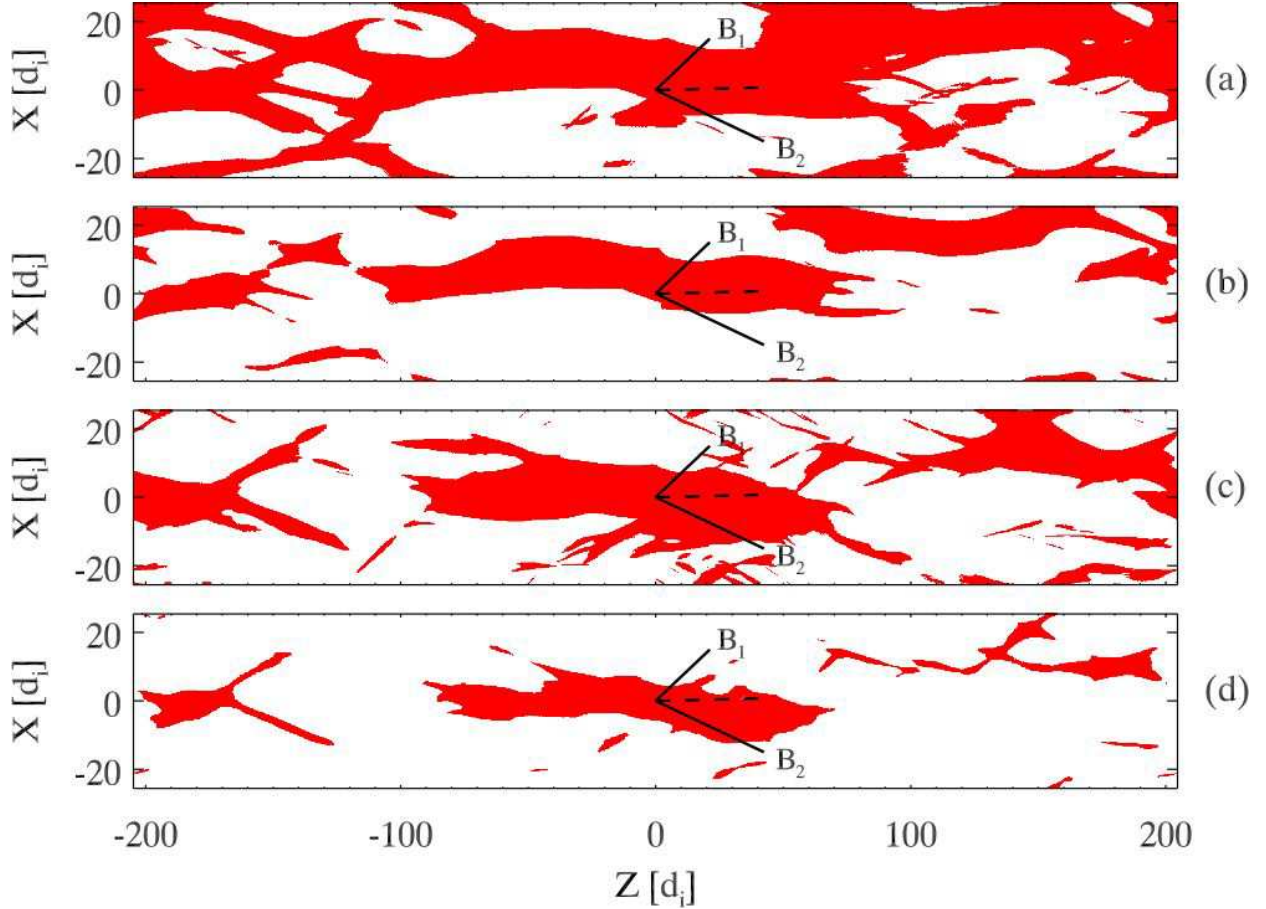}
\caption{\label{isosurface} (Color online) Top view of the isosurfaces
of $J_{xz}$ in the upper current layer at $t=84$ (panels (a) and (b))
and $t=201$ (panels (c) and (d)).  Panels (a) and (c) are at an
isosurface level of 1.3, panels (b) and (d) at 1.7.  The axes have
been shifted to put the prominent features near the center.  Solid
lines denote the asymptotic magnetic fields and the dashed line the
expected orientation of the X-line according to
\citet{swisdak07a}.  Note the different axis scales.}
\end{center}
\end{figure}

While in the 2D case topological constraints make the identification
of X-lines straightforward, in three dimensions the situation is more
complicated, particularly when, as is the case here, no 3-D nulls
exist in the system \cite{priest03a}.  In this work we take an
empirical approach to identifying reconnection sites by examining
isosurfaces of the current density $J_{xz} = \sqrt{J_x^2+J_z^2}$.  (We
use $J_{xz}$ in order to avoid favoring a particular X-line
orientation in the $x-z$ plane, although in practice we find that $J_z
\gg J_x$.)  Fig.~\ref{isosurface} shows two values of the isosurface
level at two different times.  The regions of strong current, which
map the X-lines, form extended structures in the $x$-$z$ plane at
various angles with respect to the $z$ axis.  At $t=84$ (panels (a)
and (b)) multiple X-lines are present, but by $t=201$ (panels (c) and
(d)) only a few distinct orientations dominate the system.  Although
the predominant X-line orientation at $t=201$ is horizontal, there are
some interesting features that appear to be aligned with the
asymptotic magnetic fields.  Although many of the weaker examples (for
instance, the thin structures at $z \approx 40$ and $x\approx 20$ in
panel (c)) are field-aligned currents not directly associated with
reconnection, the strongest instances (e.g., the structure at
$-170\leq z \leq -120$ and $-10 \leq x \leq 10$ in panel (d))
correspond to X-lines.  Reconnection of the initial asymptotic fields
cannot be occurring at such sites and, in fact, cuts through these
features (not shown) reveal that the local reconnecting fields differ
significantly from the initial asymptotic values.

\begin{figure}
\begin{center}
\noindent\includegraphics[width=\textwidth]{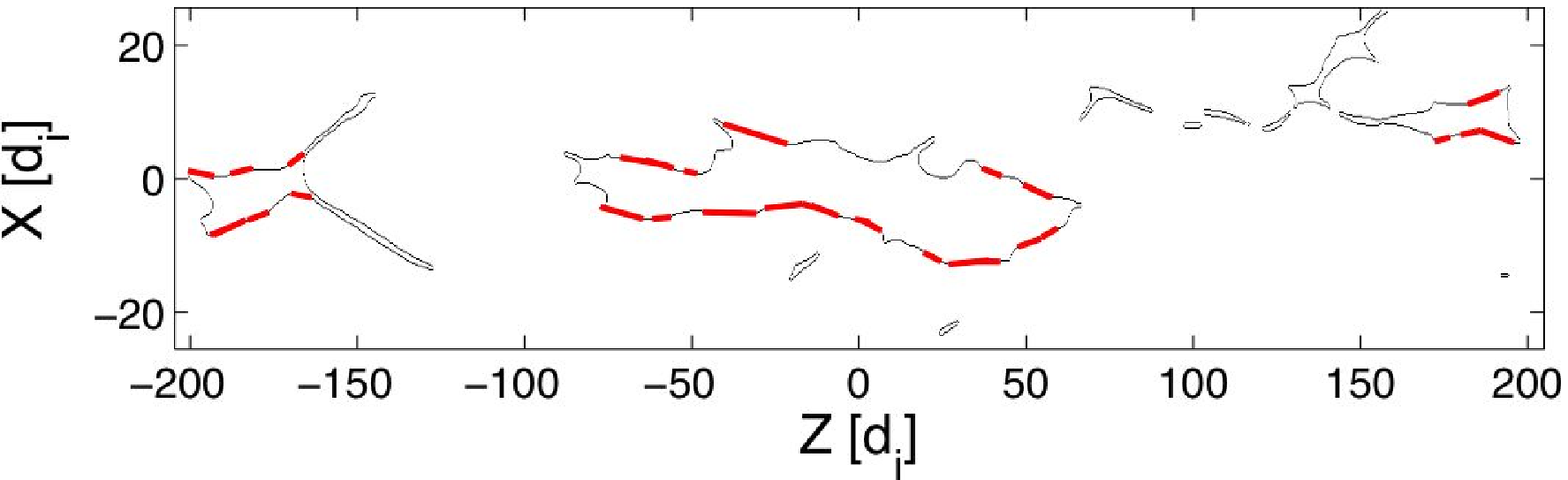}
\caption{\label{Hough} (Color online) Edges of the projection of the
strong current density at $t=201$ on the $x-z$ plane and detected by
the Canny method are shown in black.  The red lines are the result of
the Hough transform.}
\end{center}
\end{figure}

To evaluate the orientation of the X-lines quantitatively, we project
the isosurfaces on the $x-z$ plane and detect the edges with the Canny
method \citep{canny86a}, a standard image processing tool that finds
edges by looking for local maxima of the gradient. The gradient is
calculated using the derivative of a Gaussian filter.  The method uses
two thresholds to detect strong and weak edges and includes the weak
edges in the output only if they are connected to strong edges.  In
Fig.~\ref{Hough}, the edges of the isosurface projection of
Fig.~\ref{isosurface}d are shown in black.  Due to imperfections in
either the image data or the edge detector, there may be missing
points on the desired curve.

The grouping of the extracted edge features to determine the X-line
orientation is done with the Hough transform \citep{ballard81a}.  For
each image pixel and its neighborhood, the Hough transform algorithm
determines whether an edge exists at that pixel.  The pixels lying
along the highest values of parametric lines represent potential lines
in the input image.  Small gaps are automatically filled, and the
lines are identified while a threshold is applied so that only lines
longer than that value are considered.  The red lines in
Fig.~\ref{Hough} are the identified lines of the Hough transform, and
clearly map the X-line.  By simply averaging the various orientations
identified in Fig.~\ref{Hough}, we find that at $t=201$ the X-line is
oriented at an angle of $\phi\approx 0.6\pm 8.5^{\circ}$ with respect
to the $z$-axis, which corresponds to $\alpha_{\text{sim}} =
\arctan(1/\sqrt{2}) + \phi \approx 35.9^{\circ} \pm 8.5^{\circ}$.

We have demonstrated that magnetic reconnection can occur between
magnetic fields of small shear angle, in particular in configurations
in which there is no anti-parallel component of the magnetic field in
the plane perpendicular to the equilibrium current.  This result
disagrees with the model of \citet{sonnerup74a}.  In the initial stages
of the simulation, we see the growth of X-lines with multiple
orientations.  By late time, however, one direction predominates.  The
orientation of this X-line, $\alpha_{\text{sim}} = 35.9^{\circ}\pm
8.5^{\circ}$, agrees with the prediction of \citet{swisdak07a},
$\alpha_{SD} = 34.3^{\circ}$.  The existence of many short X-lines
with differing orientations early in the simulation demonstrates that
$L_z$ does not play a limiting role and hence that the size of the
computational domain probably does not affect our results.


The value of $\alpha_{SD}$ predicted by \citet{swisdak07a} is the
angle that maximizes the outflow speed from the X-line when the
reconnecting fields have their asymptotic initial values.  The strong
diagonal features in Fig.~\ref{isosurface}d discussed above, on the
other hand, are due to the reconnection of significantly perturbed
fields.  (The density contrast, $\rho_2/\rho_1$, remains essentially
constant.)  Applying the criterion of Eq. \ref{alpha} to the perturbed
fields yields orientations roughly consistent with those observed in
the simulation.  It is unclear, however, if there is some overarching
reason why the local reconnecting fields would be reconfigured in such
a way as to generate X-lines that parallel the original asymptotic
fields, or if the alignment in this case is purely coincidental.

\begin{figure}
\begin{center}
\noindent\includegraphics[width=\textwidth]{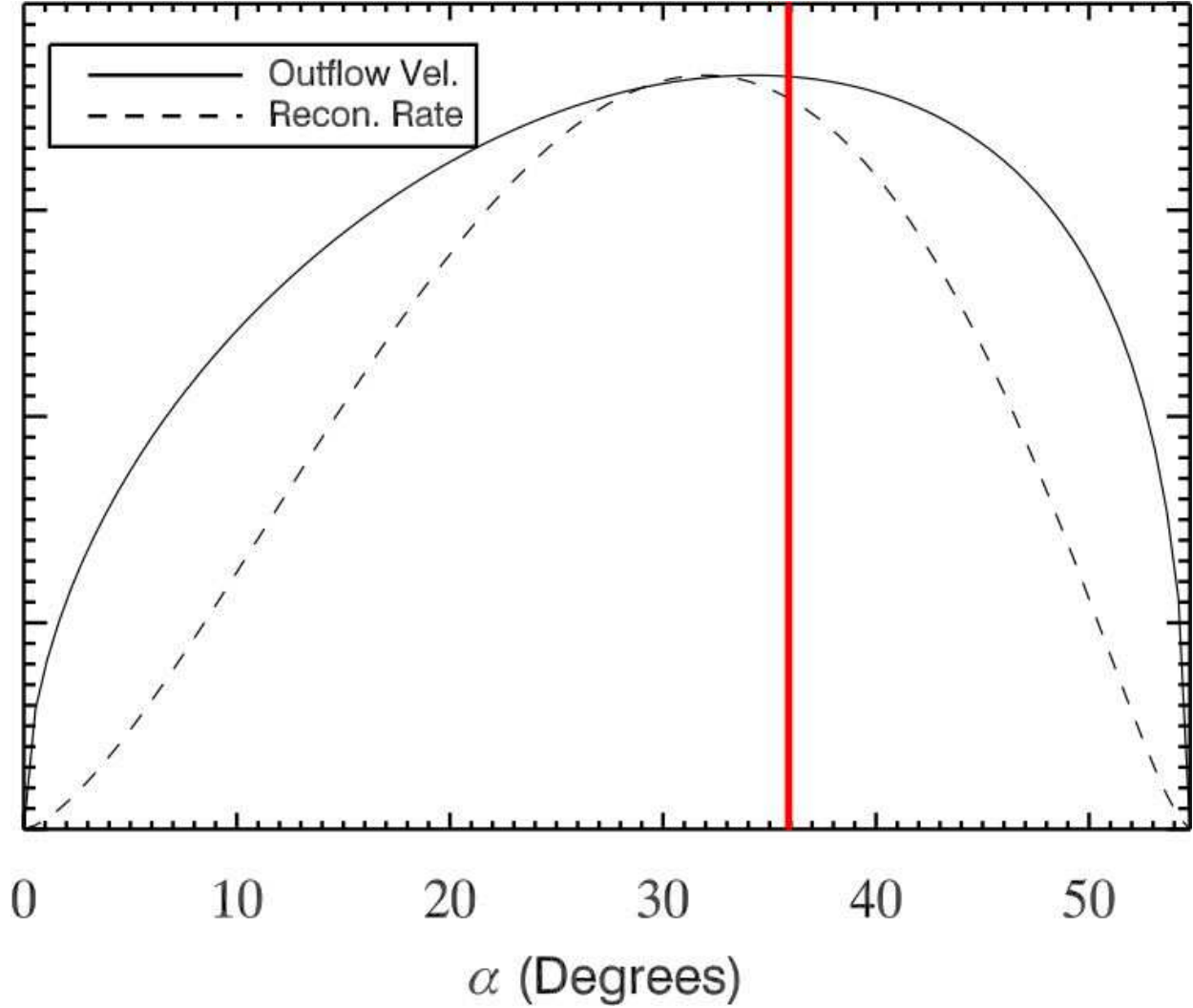}
\caption{\label{voutfig} (Color online) The magnitude of
$v_{\text{out}}$ (Equation \ref{vout}) and the normalized reconnection
rate of \citet{cassak07a} (see Equation
\ref{cs}) versus $\alpha$.  The ordinal units are arbitrary and have
been suppressed.  The vertical red line shows $\alpha_{\text{sim}}$
from our three-dimensional simulation.}
\end{center}
\end{figure}

It has been suggested (M.~A.~Shay, private communication) that
maximizing the normalized reconnection rate, and not the outflow speed
from the X-line, determines the orientation.  If the aspect ratio of
the diffusion region ($R$, assumed to be $\leq 1$) remains independent
of the upstream properties of the plasma (which has not been
established in 3-D simulations) \citet{cassak07a} argue that the
normalized rate $E$ varies as
\begin{equation}\label{cs}
E \sim 2R\,
\frac{v_{\text{out}}}{c}\left(\frac{B_{1x}B_{2x}}{B_{1x}+B_{2x}}\right)
\text{.}
\end{equation}
In Fig.~\ref{voutfig} the solid and dashed lines trace the dependence
of $v_{\text{out}}$ and $E$ on $\alpha$ for the parameters of our
simulations; the vertical red line gives $\alpha_{\text{sim}}$.  The
dashed line peaks at an angle $\alpha_{E} = 31.7^{\circ}$, which is
less than $\alpha_{SD}$ and slightly farther away from the value
measured in the simulation.  However, given the relatively broad peaks
generated by both criteria and the uncertainties associated with
determining $\alpha_{\text{sim}}$, we cannot reliably discriminate
between the two.  The similarities between the quantities being
maximized means that doing so probably requires extreme choices of
parameters (e.g., $B_2/B_1 \gg 1$) that are difficult to simulate.

As can be seen in Fig.~\ref{isosurface}, the lengths of the X-lines
barely change between $t=84$ and $t=201$; in fact most growth occurs
early in the simulation, before significant magnetic flux has
reconnected.  We find that the growth of a given X-line is usually
throttled by the interaction of its current with islands of
reconnected flux from other X-lines at different rational surfaces.
This effect will not be present in anti-parallel reconnection (where
all of the X-lines are confined to a single plane) and may explain why
our result conflicts with the finding of \citet{huba02b} that X-lines
continually grow in the current direction.  \citet{shay03a} did see
stagnation of the X-line length for some initial current sheet widths,
although not for the value used in this work ($w_0 = 0.5$).

Our results suggest that reconnection can occur in any system where
the adjoining fields are not parallel and in which other processes do
not suppress reconnection (e.g., diamagnetic drifts
\citep{swisdak03a}).  The relatively broad peak of $v_{\text{out}}$ in
Fig.~\ref{voutfig} may mean that, for a given set of asymptotic
conditions, X-lines do not take on a single orientation but instead
exhibit a distribution of orientations.  Further 3-D simulations are
needed to test this hypothesis.

We are not aware of any other model that does a better job of
predicting $\alpha$.  We suggest that on an encounter with
reconnection events in which highly asymmetric conditions exist, or
while numerically reconstructing such an event, the
\citet{swisdak07a} criterion can cautiously be applied to
determine the orientation of the reconnection X-line, as has already
been done by \citet{phan09a} and \citet{teh08a}.

\begin{acknowledgments}
R. S. would like to thank the members of the IREAP group at the
University of Maryland for their warm hospitality during his
sabbatical year.
\end{acknowledgments}


\begin{thebibliography}{18}%
\makeatletter
\providecommand \@ifxundefined [1]{%
 \@ifx{#1\undefined}
}%
\providecommand \@ifnum [1]{%
 \ifnum #1\expandafter \@firstoftwo
 \else \expandafter \@secondoftwo
 \fi
}%
\providecommand \@ifx [1]{%
 \ifx #1\expandafter \@firstoftwo
 \else \expandafter \@secondoftwo
 \fi
}%
\providecommand \natexlab [1]{#1}%
\providecommand \enquote  [1]{``#1''}%
\providecommand \bibnamefont  [1]{#1}%
\providecommand \bibfnamefont [1]{#1}%
\providecommand \citenamefont [1]{#1}%
\providecommand \href@noop [0]{\@secondoftwo}%
\providecommand \href [0]{\begingroup \@sanitize@url \@href}%
\providecommand \@href[1]{\@@startlink{#1}\@@href}%
\providecommand \@@href[1]{\endgroup#1\@@endlink}%
\providecommand \@sanitize@url [0]{\catcode `\\12\catcode `\$12\catcode
  `\&12\catcode `\#12\catcode `\^12\catcode `\_12\catcode `\%12\relax}%
\providecommand \@@startlink[1]{}%
\providecommand \@@endlink[0]{}%
\providecommand \url  [0]{\begingroup\@sanitize@url \@url }%
\providecommand \@url [1]{\endgroup\@href {#1}{\urlprefix }}%
\providecommand \urlprefix  [0]{URL }%
\providecommand \Eprint [0]{\href }%
\providecommand \doibase [0]{http://dx.doi.org/}%
\providecommand \selectlanguage [0]{\@gobble}%
\providecommand \bibinfo  [0]{\@secondoftwo}%
\providecommand \bibfield  [0]{\@secondoftwo}%
\providecommand \translation [1]{[#1]}%
\providecommand \BibitemOpen [0]{}%
\providecommand \bibitemStop [0]{}%
\providecommand \bibitemNoStop [0]{.\EOS\space}%
\providecommand \EOS [0]{\spacefactor3000\relax}%
\providecommand \BibitemShut  [1]{\csname bibitem#1\endcsname}%
\let\auto@bib@innerbib\@empty
\bibitem [{\citenamefont {Sonnerup}(1974)}]{sonnerup74a}%
  \BibitemOpen
  \bibfield  {author} {\bibinfo {author} {\bibfnamefont {B.~U.~{\"O}.}\
  \bibnamefont {Sonnerup}},\ }\href@noop {} {\bibfield  {journal} {\bibinfo
  {journal} {J. Geophys. Res.}\ }\textbf {\bibinfo {volume} {79}},\ \bibinfo
  {pages} {1546} (\bibinfo {year} {1974})}\BibitemShut {NoStop}%
\bibitem [{\citenamefont {Cowley}(1976)}]{cowley76a}%
  \BibitemOpen
  \bibfield  {author} {\bibinfo {author} {\bibfnamefont {S.~W.~H.}\
  \bibnamefont {Cowley}},\ }\href@noop {} {\bibfield  {journal} {\bibinfo
  {journal} {J. Geophys. Res.}\ }\textbf {\bibinfo {volume} {81}},\ \bibinfo
  {pages} {3455} (\bibinfo {year} {1976})}\BibitemShut {NoStop}%
\bibitem [{\citenamefont {Teh}\ and\ \citenamefont {Sonnerup}(2008)}]{teh08a}%
  \BibitemOpen
  \bibfield  {author} {\bibinfo {author} {\bibfnamefont {W.-L.}\ \bibnamefont
  {Teh}}\ and\ \bibinfo {author} {\bibfnamefont {B.~U.~{\"O}.}\ \bibnamefont
  {Sonnerup}},\ }\href@noop {} {\bibfield  {journal} {\bibinfo  {journal} {Ann.
  Geophys.}\ }\textbf {\bibinfo {volume} {26}},\ \bibinfo {pages} {2673}
  (\bibinfo {year} {2008})}\BibitemShut {NoStop}%
\bibitem [{\citenamefont {Gosling}(2007)}]{gosling07a}%
  \BibitemOpen
  \bibfield  {author} {\bibinfo {author} {\bibfnamefont {J.~T.}\ \bibnamefont
  {Gosling}},\ }\href@noop {} {\bibfield  {journal} {\bibinfo  {journal} {Ap.
  J. Lett.}\ }\textbf {\bibinfo {volume} {671}},\ \bibinfo {pages} {L73}
  (\bibinfo {year} {2007})}\BibitemShut {NoStop}%
\bibitem [{\citenamefont {Gosling}\ \emph {et~al.}(2007)\citenamefont
  {Gosling}, \citenamefont {Phan}, \citenamefont {Lin},\ and\ \citenamefont
  {Szabo}}]{gosling07b}%
  \BibitemOpen
  \bibfield  {author} {\bibinfo {author} {\bibfnamefont {J.~T.}\ \bibnamefont
  {Gosling}}, \bibinfo {author} {\bibfnamefont {T.~D.}\ \bibnamefont {Phan}},
  \bibinfo {author} {\bibfnamefont {R.~P.}\ \bibnamefont {Lin}}, \ and\
  \bibinfo {author} {\bibfnamefont {A.}~\bibnamefont {Szabo}},\ }\href
  {\doibase 10.1029/2007GL030706} {\bibfield  {journal} {\bibinfo  {journal}
  {Geophys. Res. Lett.}\ }\textbf {\bibinfo {volume} {34}},\ \bibinfo {eid}
  {L15110} (\bibinfo {year} {2007}),\ 10.1029/2007GL030706}\BibitemShut
  {NoStop}%
\bibitem [{\citenamefont {Phan}, \citenamefont {Gosling},\ and\ \citenamefont
  {Davis}(2009)}]{phan09a}%
  \BibitemOpen
  \bibfield  {author} {\bibinfo {author} {\bibfnamefont {T.~D.}\ \bibnamefont
  {Phan}}, \bibinfo {author} {\bibfnamefont {J.~T.}\ \bibnamefont {Gosling}}, \
  and\ \bibinfo {author} {\bibfnamefont {M.~S.}\ \bibnamefont {Davis}},\ }\href
  {\doibase 10.1029/2009GL037713} {\bibfield  {journal} {\bibinfo  {journal}
  {Geophys. Res. Lett.}\ }\textbf {\bibinfo {volume} {36}},\ \bibinfo {eid}
  {L09108} (\bibinfo {year} {2009}),\ 10.1029/2009GL037713}\BibitemShut
  {NoStop}%
\bibitem [{\citenamefont {Phan}\ \emph {et~al.}(2010)\citenamefont {Phan},
  \citenamefont {Gosling}, \citenamefont {Paschmann}, \citenamefont {Pasma},
  \citenamefont {Drake}, \citenamefont {{\O}ieroset}, \citenamefont {Larson},
  \citenamefont {Lin},\ and\ \citenamefont {Davis}}]{phan10a}%
  \BibitemOpen
  \bibfield  {author} {\bibinfo {author} {\bibfnamefont {T.~D.}\ \bibnamefont
  {Phan}}, \bibinfo {author} {\bibfnamefont {J.~T.}\ \bibnamefont {Gosling}},
  \bibinfo {author} {\bibfnamefont {G.}~\bibnamefont {Paschmann}}, \bibinfo
  {author} {\bibfnamefont {C.}~\bibnamefont {Pasma}}, \bibinfo {author}
  {\bibfnamefont {J.~F.}\ \bibnamefont {Drake}}, \bibinfo {author}
  {\bibfnamefont {M.}~\bibnamefont {{\O}ieroset}}, \bibinfo {author}
  {\bibfnamefont {D.}~\bibnamefont {Larson}}, \bibinfo {author} {\bibfnamefont
  {R.~P.}\ \bibnamefont {Lin}}, \ and\ \bibinfo {author} {\bibfnamefont
  {M.~S.}\ \bibnamefont {Davis}},\ }\href {\doibase 10.1088/2041-8205/719/L199}
  {\bibfield  {journal} {\bibinfo  {journal} {Astrophys. J.. Lett.}\ }\textbf
  {\bibinfo {volume} {719}},\ \bibinfo {pages} {L199} (\bibinfo {year}
  {2010})}\BibitemShut {NoStop}%
\bibitem [{\citenamefont {Swisdak}\ and\ \citenamefont
  {Drake}(2007)}]{swisdak07a}%
  \BibitemOpen
  \bibfield  {author} {\bibinfo {author} {\bibfnamefont {M.}~\bibnamefont
  {Swisdak}}\ and\ \bibinfo {author} {\bibfnamefont {J.~F.}\ \bibnamefont
  {Drake}},\ }\href {\doibase 10.1029/2007GL029815} {\bibfield  {journal}
  {\bibinfo  {journal} {Geophys. Res. Lett.}\ }\textbf {\bibinfo {volume}
  {34}},\ \bibinfo {eid} {L11106} (\bibinfo {year} {2007}),\
  10.1029/2007GL029815}\BibitemShut {NoStop}%
\bibitem [{\citenamefont {Cassak}\ and\ \citenamefont
  {Shay}(2007)}]{cassak07a}%
  \BibitemOpen
  \bibfield  {author} {\bibinfo {author} {\bibfnamefont {P.~A.}\ \bibnamefont
  {Cassak}}\ and\ \bibinfo {author} {\bibfnamefont {M.~A.}\ \bibnamefont
  {Shay}},\ }\href {\doibase 10.1063/1.2795630} {\bibfield  {journal} {\bibinfo
   {journal} {Phys. Plasmas}\ }\textbf {\bibinfo {volume} {14}},\ \bibinfo
  {eid} {102114} (\bibinfo {year} {2007}),\ 10.1063/1.2795630}\BibitemShut
  {NoStop}%
\bibitem [{\citenamefont {Huba}\ and\ \citenamefont {Rudakov}(2002)}]{huba02b}%
  \BibitemOpen
  \bibfield  {author} {\bibinfo {author} {\bibfnamefont {J.~D.}\ \bibnamefont
  {Huba}}\ and\ \bibinfo {author} {\bibfnamefont {L.~I.}\ \bibnamefont
  {Rudakov}},\ }\href@noop {} {\bibfield  {journal} {\bibinfo  {journal} {Phys.
  Plasmas}\ }\textbf {\bibinfo {volume} {9}},\ \bibinfo {pages} {4435}
  (\bibinfo {year} {2002})}\BibitemShut {NoStop}%
\bibitem [{\citenamefont {Shay}\ \emph {et~al.}(2003)\citenamefont {Shay},
  \citenamefont {Drake}, \citenamefont {Swisdak}, \citenamefont {Dorland},\
  and\ \citenamefont {Rogers}}]{shay03a}%
  \BibitemOpen
  \bibfield  {author} {\bibinfo {author} {\bibfnamefont {M.~A.}\ \bibnamefont
  {Shay}}, \bibinfo {author} {\bibfnamefont {J.~F.}\ \bibnamefont {Drake}},
  \bibinfo {author} {\bibfnamefont {M.}~\bibnamefont {Swisdak}}, \bibinfo
  {author} {\bibfnamefont {W.}~\bibnamefont {Dorland}}, \ and\ \bibinfo
  {author} {\bibfnamefont {B.~N.}\ \bibnamefont {Rogers}},\ }\href {\doibase
  10.1029/2002GL016267} {\bibfield  {journal} {\bibinfo  {journal} {Geophys.
  Res. Lett.}\ }\textbf {\bibinfo {volume} {30}} (\bibinfo {year} {2003}),\
  10.1029/2002GL016267}\BibitemShut {NoStop}%
\bibitem [{\citenamefont {Lapenta}\ \emph {et~al.}(2006)\citenamefont
  {Lapenta}, \citenamefont {Krauss-Varban}, \citenamefont {Karimabadi},\ and\
  \citenamefont {Huba}}]{lapenta06a}%
  \BibitemOpen
  \bibfield  {author} {\bibinfo {author} {\bibfnamefont {G.}~\bibnamefont
  {Lapenta}}, \bibinfo {author} {\bibfnamefont {D.}~\bibnamefont
  {Krauss-Varban}}, \bibinfo {author} {\bibfnamefont {H.}~\bibnamefont
  {Karimabadi}}, \ and\ \bibinfo {author} {\bibfnamefont {J.~D.}\ \bibnamefont
  {Huba}},\ }\href {\doibase 10.1029/2005GL025124} {\bibfield  {journal}
  {\bibinfo  {journal} {Geophys. Res. Lett.}\ }\textbf {\bibinfo {volume}
  {33}},\ \bibinfo {eid} {L10102} (\bibinfo {year} {2006}),\
  10.1029/2005GL025124}\BibitemShut {NoStop}%
\bibitem [{\citenamefont {Shay}\ \emph {et~al.}(2004)\citenamefont {Shay},
  \citenamefont {Drake}, \citenamefont {Swisdak},\ and\ \citenamefont
  {Rogers}}]{shay04a}%
  \BibitemOpen
  \bibfield  {author} {\bibinfo {author} {\bibfnamefont {M.~A.}\ \bibnamefont
  {Shay}}, \bibinfo {author} {\bibfnamefont {J.~F.}\ \bibnamefont {Drake}},
  \bibinfo {author} {\bibfnamefont {M.}~\bibnamefont {Swisdak}}, \ and\
  \bibinfo {author} {\bibfnamefont {B.~N.}\ \bibnamefont {Rogers}},\
  }\href@noop {} {\bibfield  {journal} {\bibinfo  {journal} {Phys. Plasmas}\
  }\textbf {\bibinfo {volume} {11}},\ \bibinfo {pages} {2199} (\bibinfo {year}
  {2004})}\BibitemShut {NoStop}%
\bibitem [{\citenamefont {Guzdar}\ \emph {et~al.}(1993)\citenamefont {Guzdar},
  \citenamefont {Drake}, \citenamefont {McCrathy}, \citenamefont {Hassam},\
  and\ \citenamefont {Liu}}]{guzdar93a}%
  \BibitemOpen
  \bibfield  {author} {\bibinfo {author} {\bibfnamefont {P.~N.}\ \bibnamefont
  {Guzdar}}, \bibinfo {author} {\bibfnamefont {J.~F.}\ \bibnamefont {Drake}},
  \bibinfo {author} {\bibfnamefont {D.}~\bibnamefont {McCrathy}}, \bibinfo
  {author} {\bibfnamefont {A.~B.}\ \bibnamefont {Hassam}}, \ and\ \bibinfo
  {author} {\bibfnamefont {C.~S.}\ \bibnamefont {Liu}},\ }\href@noop {}
  {\bibfield  {journal} {\bibinfo  {journal} {Phys. Fluids B}\ }\textbf
  {\bibinfo {volume} {5}},\ \bibinfo {pages} {3712} (\bibinfo {year}
  {1993})}\BibitemShut {NoStop}%
\bibitem [{\citenamefont {Priest}, \citenamefont {Hornig},\ and\ \citenamefont
  {Pontin}(2003)}]{priest03a}%
  \BibitemOpen
  \bibfield  {author} {\bibinfo {author} {\bibfnamefont {E.~R.}\ \bibnamefont
  {Priest}}, \bibinfo {author} {\bibfnamefont {G.}~\bibnamefont {Hornig}}, \
  and\ \bibinfo {author} {\bibfnamefont {D.~I.}\ \bibnamefont {Pontin}},\
  }\href {\doibase 10.1029/2002JA009812} {\bibfield  {journal} {\bibinfo
  {journal} {J. Geophys. Res.}\ }\textbf {\bibinfo {volume} {108}},\ \bibinfo
  {pages} {1285} (\bibinfo {year} {2003})}\BibitemShut {NoStop}%
\bibitem [{\citenamefont {Canny}(1986)}]{canny86a}%
  \BibitemOpen
  \bibfield  {author} {\bibinfo {author} {\bibfnamefont {J.~A.}\ \bibnamefont
  {Canny}},\ }\href@noop {} {\bibfield  {journal} {\bibinfo  {journal} {{IEEE}
  Trans., Pattern Analysis and Machine Intelligence}\ }\textbf {\bibinfo
  {volume} {8}},\ \bibinfo {pages} {679} (\bibinfo {year} {1986})}\BibitemShut
  {NoStop}%
\bibitem [{\citenamefont {Ballard}(1981)}]{ballard81a}%
  \BibitemOpen
  \bibfield  {author} {\bibinfo {author} {\bibfnamefont {D.~H.}\ \bibnamefont
  {Ballard}},\ }\href@noop {} {\bibfield  {journal} {\bibinfo  {journal}
  {Pattern Recognition}\ }\textbf {\bibinfo {volume} {13}},\ \bibinfo {pages}
  {111} (\bibinfo {year} {1981})}\BibitemShut {NoStop}%
\bibitem [{\citenamefont {Swisdak}\ \emph {et~al.}(2003)\citenamefont
  {Swisdak}, \citenamefont {Rogers}, \citenamefont {Drake},\ and\ \citenamefont
  {Shay}}]{swisdak03a}%
  \BibitemOpen
  \bibfield  {author} {\bibinfo {author} {\bibfnamefont {M.}~\bibnamefont
  {Swisdak}}, \bibinfo {author} {\bibfnamefont {B.~N.}\ \bibnamefont {Rogers}},
  \bibinfo {author} {\bibfnamefont {J.~F.}\ \bibnamefont {Drake}}, \ and\
  \bibinfo {author} {\bibfnamefont {M.~A.}\ \bibnamefont {Shay}},\ }\href
  {\doibase 10.1029/2002JA009726} {\bibfield  {journal} {\bibinfo  {journal}
  {J. Geophys. Res.}\ }\textbf {\bibinfo {volume} {108}},\ \bibinfo {pages}
  {1218} (\bibinfo {year} {2003})}\BibitemShut {NoStop}%
\end{thebibliography}
%

\end{document}